\documentclass[aps, prd, twocolumn, superscriptaddress, nofootinbib,
amsmath, amssymb]{revtex4-1}

\usepackage{graphicx}
\usepackage{mathrsfs} 
\usepackage[normalem]{ulem} 

\newcommand{\Tr}{\ensuremath{\mathrm{Tr}}}

\begin{document}

\title{Spectrum of Trace Deformed Yang-Mills Theories}

\author{Andreas Athenodorou}
\email{andreas.athinodorou@pi.infn.it}
\author{Marco Cardinali}
\email{marco.cardinali@pi.infn.it}
\author{and Massimo D'Elia}
\email{massimo.delia@unipi.it}


\affiliation{Dipartimento di Fisica, Universit\`a di Pisa and INFN, Sezione di Pisa, Largo Pontecorvo 3, 56127 Pisa, Italy}


\begin{abstract}
In this paper we study, by means of numerical simulations, the behaviour of the scalar glueball mass and the ground state of the torelon for trace deformed Yang-Mills 
theory defined on $ \mathbb{R}^3\times S^1$, in which center symmetry is recovered even at small compactification radii. 
We find, by investigating both $SU(3)$ and $SU(4)$ pure gauge theories, that the glueball mass computed in the deformed theory, when center symmetry is recovered, is compatible with its value at zero temperature and does not show any significant dependence on the compactification radius; moreover, we establish a connection between the deformation parameter
and an effective compactification size, which works well at least for small deformations.  
In addition, we observe that the ground state of the torelon which winds around the small traced deformed circle with size $l$ acquires a pleateau for large values of the strength $h$, with values which are compatible with 
a $1/l$ behavior but, on the other hand, are still not in complete agreement
with the asymptotic semiclassical 
large-$N$ predictions.
%
\end{abstract}

\maketitle

\section{Introduction}  

\label{sec:intro}

Large-$N$ volume independence of $SU(N)$ Yang-Mills 
and similar theories is a topic 
discussed since 
long~\cite{Eguchi:1982nm,Yaffe:1981vf,Bhanot:1982sh,GonzalezArroyo:1982hz,Kovtun:2007py,Unsal:2010qh}. The possibility that the non-perturbative 
properties of the theory might be encoded in a simplified model with 
small compactification radii is very appealing, also in view of the 
fact that the inverse compactification radius sets a high energy 
scale, which makes weak coupling approaches viable. It has been clarified 
since long that volume independence holds only 
when no transition, leading to the breaking of 
center symmetry, takes place:
unfortunately, the breaking happens in most cases, e.g., at the thermal
radius where the theory deconfines.

More recently, trace deformed theories~\cite{Unsal:2008ch,Myers:2007vc} 
have been introduced, as a way to maintain center symmetry unbroken
even in the presence of arbitrarily small compactification radii.
That provides a tool to study the issue of volume independence
in a controlled way, e.g. by numerical lattice simulations,
and also to compare with the expectations from 
semiclassical analytic computations.
The idea, inspired by the perturbative form of the Polyakov loop
effective action at high $T$~\cite{Gross:1980br}, 
is to introduce center symmetric couplings to the Polyakov loop and its powers,
so as to inhibit the spontaneous breaking of center symmetry.
This offers the possibility to test volume independence 
and, at the same time, to investigate the connection of 
center symmetry to many other non-perturbative features of 
Yang-Mills theories.

Several studies have already considered the use of trace deformed theories
and also of possible alternatives, like the introduction of adjoint 
fermions~\cite{Kovtun:2007py, Unsal:2007vu, Unsal:2007jx, Shifman:2008ja, Myers:2009df,
Cossu:2009sq, Meisinger:2009ne, Unsal:2010qh, Thomas:2012ib, Poppitz:2012sw,
Thomas:2012tu, Poppitz:2012nz, Misumi:2014raa, Anber:2014lba, Bhoonah:2014gpa,
Cherman:2016vpt, Sulejmanpasic:2016llc, Anber:2017rch}. 
From the point of view of lattice simulations, recent 
studies have provided extensive numerical evidence regarding
$\theta$-dependence in $SU(3)$~\cite{Bonati:2018rfg}
and $SU(4)$~\cite{Bonati:2019kmf} gauge theories,
considering in particular 
the first two coefficients of the Taylor expansion
of the free energy expansion as a function of $\theta$.
The remarkable result is that, even for compactification
radii as small as $(500\, \textrm{MeV})^{-1}$, one recovers, within
numerical errors, the same $\theta$-dependence
as in the confined phase as soon as center symmetry is completely 
restored by means of the trace deformations.
This goes even beyond the expectations from analytic semiclassical 
computations, since the volume independence, 
at least for what concerns $\theta$-dependence,
is quantitatively exact (within errors) and observed 
for not-too-large values of $N$. The result is even more 
interesting when considering that, from a dynamical point of view,
the restoration of center symmetry takes place in a slightly different
way in the standard confined phase and in the deformed 
theory~\cite{Bonati:2018rfg, Bonati:2019kmf}, so it seems that it is 
just the realization of the symmetry that counts,
at least for $\theta$-dependence, independently of other details.

That claims for a deeper investigation, regarding also 
other non-perturbative properties of Yang-Mills theories.
In this study we consider the physical spectrum of the theory,
which for a pure gauge theory consists mostly of glueballs.
The glueball spectrum is known to undergo strong modifications
when crossing the deconfinement transition~\cite{Falcone:2007rv,Grossman:1993wm}, therefore 
checking if it goes back to the $T = 0$ spectrum by just 
recovering center symmetry represents a strong test of volume independence:
anticipating somewhat our conclusions and the final discussion, the answer
is positive, at least for the $SU(3)$ and $SU(4)$ gauge theories which are
investigated in the present study. 
A different issue regards the torelon masses, i.e.~the masses of physical 
states corresponding to flux tubes winding around the compactified 
direction: this has no correspondence with the infinite volume theory,
however there are well definite semiclassical 
predictions~\cite{Unsal:2008ch} that one would like to compare with.

The paper is organized as follows. In Section~\ref{sec:setup}
we summarize the definition of $SU(N)$ pure gauge theories 
in the presence of trace deformations and our lattice implementation.
In Section~\ref{sec:mass} we review the theoretical and numerical
methods adopted to determine the glueball and torelon
spectrum in our numerical simulations.
In Section~\ref{sec:results} we present our numerical results for
$SU(3)$,
which consist of determinations of glueball and torelon masses
both with and without trace deformations, in order to 
assess a possible correspondence between the two cases, and for different
values of the lattice spacing, in order to assess
the relevance of cutoff effects. In Section~\ref{sec:su4} we present
some exploratory results regarding the $SU(4)$ case.
Finally, in Section~\ref{sec:conclusion}, we draw our conclusions. 

\section{Numerical Setup}

\label{sec:setup}

\subsection{Traced deformed Yang-Mills theory}

{Trace deformed $SU(N)$ Yang-Mills (YM) theories were first proposed in Ref.~\cite{Unsal:2008ch}, although a previous lattice study was already presented in Ref.~\cite{Myers:2007vc}. Additional center symmetric couplings to the Polyakov loop are added to the usual YM action, in order to prevent center symmetry  breaking, even for compactification length $l$ smaller than the critical one $l_c$. The action of trace deformed $SU(N)$ YM theory is thus given by~\cite{Unsal:2008ch}}:
\begin{equation}\label{deformed_action}
S^{\mathrm{def}} = S_{YM} + \sum_{\vec{n}} \sum_{j=1}^{\lfloor N/2\rfloor} h_j |\Tr P^{j}(\vec n)|^2 \ ,
\end{equation}
{where $\vec{n}$ identifies a point in the space orthogonal to the compactified direction, 
$\lfloor \quad \rfloor$ denotes the floor function, $P$ is the Polyakov loop in the compactified 
direction and the $h_j$s are the deformation parameters, 
corresponding to the number of independent, center-symmetric functions of the 
Polyakov loop; 1 and 2 parameters are needed 
respectively for the cases $N = 3$ and $N = 4$ explored in the present study.} 
$S_{\rm YM}$ is the standard YM plaquette action:
\begin{equation}
 S_{\rm YM} = \beta \sum_{\square} \left\{1-\frac{1}{N} {\text{ReTr}} U_{\square}\right\}  
\quad ; \quad \beta=\frac{2N}{g^2}.
\label{eqn_S}
\end{equation}
%
Trace deformation and other alternatives (like the introduction of adjoint fermions or the
use of non-thermal boundary conditions) have already been studied in several previous works~\cite{Kovtun:2007py, Unsal:2007vu, Unsal:2007jx, Shifman:2008ja, Myers:2009df, Cossu:2009sq,  Meisinger:2009ne, Unsal:2010qh, Thomas:2012ib, Poppitz:2012sw, Thomas:2012tu, Poppitz:2012nz, Misumi:2014raa, Anber:2014lba, Bhoonah:2014gpa, Cherman:2016vpt, Aitken:2017ayq, Sulejmanpasic:2016llc, Anber:2017rch, Tanizaki:2019rbk, Itou:2018wkm, Bergner:2018unx}. {The possibility of preserving center symmetry, also in the limit of vanishing compactification radius, is quite intriguing; weak coupling methods based on the small compactified direction could be used to study the confining properties of YM and the volume independence predicted by Eguchi and Kawai in Ref.~\cite{Eguchi:1982nm} could be exploited. Moreover, trace deformation is also a powerful tool to investigate the connection between the realization of center symmetry and the properties typical of the low temperature, confining region. In} Refs.~\cite{Bonati:2018rfg, Bonati:2019kmf} {the $\theta$ dependence of the phase in which center symmetry is recovered has been studied both 
for $SU(3)$ and $SU(4)$. In both cases topological observables reach a plateau as soon as the deformation coupling $h$ is high enough to restore full center symmetry. The plateau value of such observables is also compatible with the corresponding value of the undeformed theory. This result shows, at least for topological observables, that the center stabilized phase 
has the same non-perturbative properties of the usual confining, 
zero temperature one.

\subsection{Mass extraction on the lattice}
\label{sec:mass}
Masses of colour singlet states in lattice gauge theories can be calculated using the standard decomposition of a Euclidean correlator of some operator $\phi(t)$, with high enough overlap onto the physical states in terms of the energy eigenstates of the Hamiltonian $H$:
\begin{eqnarray}
\langle \phi^\dagger(t=an_t)\phi(0) \rangle
& = &
\langle \phi^\dagger e^{-Han_t} \phi \rangle
=
\sum_i |c_i|^2 e^{-aE_in_t} \nonumber \\
& \stackrel{t\to \infty}{=} & 
|c_0|^2 e^{-aE_0n_t},
\label{extract_mass}
\end{eqnarray}
where the energy levels are ordered, $E_{i+1}\geq E_i$, with  $E_0$ that of the ground state. The only states that contribute in the above summation are those that have non zero overlaps i.e. $c_i = \langle {\rm vac} | \phi^\dagger | i \rangle \neq 0$. We, therefore, need to match the quantum numbers of the operator $\phi$ to those of the state we are interested in. In this work we are interested in glueballs and torelons, thus, we need to encode the right quantum properties within the operator $\phi$ which will enable us to project onto the aforementioned states.

The extraction of the ground state relies on how good the overlap is onto this state and how fast in $t$ we obtain the exponential decay according to Eq.~(\ref{extract_mass}). The overlap can be maximized by building operator(s) which "capture" the right properties of the state, in other words by projecting onto the right quantum numbers as well as onto the physical length scales of the relevant state. In order to achieve a decay behaviour setting in at low values of $t$ one has to minimize contributions from excited states. To this purpose we employ the variational calculation or GEVP (Generalized Eigenvalue Problem)~\cite{Luscher:1984is} applied to a basis of operators built by the same lattice path albeit in several blocking levels \cite{MT-block,BLMTUW-2004}. This reduces the contamination of excitation states onto the ground state and maximizes the overlap of the operators onto the physical length scales.    

\subsubsection{Extracting the scalar glueball mass}
\label{sec:glueball_mass}

We extract the ground state mass of the scalar glueball by making use of the variational calculation. To this purpose we employ two different operators for accomplishing the GEVP. Namely, we use the simple plaquette operator as well as the rectangular operators with size $1 a \times 2 a$ and $2 a \times 1 a$. We take linear combinations of such operators along perpendicular spatial slices so that the resulting operator has the right $0^{+ +}$ rotational properties. 

On a homogeneous cubic spatial lattice with all the spatial sizes being equal and the action being homogeneous along all spatial directions the rotational symmetry is described by the octahedral subgroup of the full rotation group. There are only five irreducible representations within this group usually labelled as $R= A_1$, $A_2$, $E$, $T_1$, $T_2$. These five irreducible representations have dimensions  of 1, 1, 2, 3, 3 respectively. The states one can calculate will belong to these five representations. In addition, glueball states are characterised by the discrete quantum numbers of Parity $P$ and Charge Conjugation $C$. Hence, at finite lattice spacing the glueball states will be labeled by $R^{PC}$. As the lattice spacing tends to zero 
one recovers the full rotational invariance, with the states falling into the $2J+ 1$ multiplets labelled by the value of angular momentum $J$. So, in principle, the glueball states can be characterized by the angular momentum $J$. The ground state characterized by $A_1^{++}$ identifies that described by quantum numbers $J^{PC}=0^{++}$.
Therefore, we will focus on the $A_1^{++}$ representation, 
assuming that its ground state provides the 
scalar ground state glueball mass.

If we choose the length in the $x$ direction to be smaller than the other two spatial sizes and/or switch on the trace deformation, i.e.~$h \neq 0$, the rotational symmetry cannot characterise the states irreducibly any more and the variational calculation built for the $A_1$ representation projects onto all irreducible representations of the octahedral group of rotations. Nevertheless, the ground state of the calculation still belongs to the $A_1^{++}$ channel and, thus, reflects the scalar glueball mass. Since operators can exchange intermediate glueballs along the toroidally compactified boundaries we should also be aware of finite volume effects along the $x$ direction. Recent calculations of the glueball spectrum~\cite{Athenodorou:2020ani,Athenodorou:2021qvs} provide bounds below which one would expect to experience such effects. 

\subsubsection{Extracting the torelon mass}
\label{sec:torelon_mass}
In the same manner we can extract the mass of the torelon which winds around the compactified deformed direction.
We obtain the torelon mass by calculating the ground state energy $m_T(h,L)$ of a flux tube of length $L$ that closes on itself by winding once around the spatial compactified deformed torus. We use Eq.~(\ref{extract_mass}) where the operator $\phi$
is the product of $SU(N)$ link matrices taken around a non-contractible closed path that winds once around the spatial torus. The simplest such operator is the elementary Polyakov loop:
\begin{equation}
\phi(n_t) = P(n_t) =  \sum_{n_y,n_z} \mathrm{Tr} 
\left\{\prod^{L_x}_{n_x=1} U_x(n_x,n_y,n_z,n_t)\right\} \,.
\label{eqn_poly}
\end{equation}
The above formula denotes the path ordered product of link matrices in the $x$-direction winding once around the $x$-torus. Then we sum over translations along the $x$-torus and a time slice so that we
project onto zero longitudinal as well as transverse momentum respectively i.e. $(p_x,p_y,p_z) = (0,0,0)$. 

The above operator is invariant under rotations about its torelon axis and so has angular momentum $J=0$. It is also clearly invariant under a combined parity and charge conjugation transformation ($CP$). Therefore, this operator is ideal for projecting onto the torelon ground state. Once more we employ smearing and blocking techniques in order to enhance the projection onto the physical states. Torelon operators are not invariant under center symmetry transformations 
and, thus, their vacuum expectation value is zero as long as this is 
not broken.

\section{Numerical results for $SU(3)$}
\label{sec:results}

The main goal of this paper is to investigate how the masses of glueballs and torelons behave in the presence of a trace deformed term in the action along the $x$-circle. For this reason, in most of our simulations we consider lattices 
with a small fixed extent along the $x$-direction (while the rest of the lattice sizes are kept fixed 
to a much larger value), such that the undeformed theory stays in the deconfined phase.
By switching on the trace deformations 
we begin to inhibit the spontaneous breaking of center symmetry:
as expected, above some threshold, confinement is restored and one might expect 
the glueball mass to acquire a value consistent with that for $T=0$. 

However, before going to the main point, we will present an extensive study of the 
spectrum in the non-deformed theory. The reason is that 
an alternative way to recover confinement, even in absence of any deformation, 
is by increasing the lattice size in the $x$-direction: once the length $l_x$ is larger than the 
deconfining critical length $l_c$, confinement is restored.
Therefore, investigating how the glueball and torelon spectrum behave as a function of $l_x$ 
will set the ground for a meaningful comparison with the behavior of the spectrum at fixed $l_x$ as a function
of the deformation parameter: that will permit to interpret results at 
non-zero $h$ in terms of an effective $l_x$ and will shed light on some of our findings.

\subsection{Results for the non-deformed theory}
\label{sec:non-deformed_theory}

We discuss glueball masses at first, then torelons. The latter present a non-trivial dependence 
on the compactification radius even in the confined phase, which is relevant to the discussion
of results obtained for the deformed theory.

\subsubsection{Glueballs in the non-deformed theory}
\label{sec:glueballs_non_deformed}
We begin our study by performing an exploratory investigation of the glueball ground state mass. To understand how the ground state mass behaves for $h=0$ and a lattice volume with one of the lattice lengths smaller than the others, we perform a quick calculation for a sequence of increasing $L_x$ while keeping the other three lattice extents fixed to $L_y= L_z= L_t =30$. In Fig.~\ref{fig:glueball_lx} we present the ground state of $A_1^{++}$ at $\beta=6.0$ which results from the GEVP using the simple plaquette as well as the rectangular $1 \times 2$ and $2 \times 1$ operators in 5 different blocking levels. The main features of this figure can be described in three different regimes of length $L_x$. 

Starting from the largest value of $L_x$ and moving downwards, at $L_x=30$ the mass of our variational calculation is simply the mass of the ground state of the scalar glueball with quantum numbers $A_1^{++}$. The hypercubic lattice is homogeneous and therefore there are no mixings with other irreducible representations of the hypercubic group of rotations. As we start reducing the length in the $x$ direction, the variational calculation although starts projecting onto all irreducible representations of the octahedral group of rotations, its ground state is still that of the $A_1^{++}$ channel. As explained before, due to the exchange of intermediate glueballs along the toroidally 
compactified boundary, we should be aware of finite volume effects along the $x$-direction. Recent calculations of the glueball spectrum~\cite{Athenodorou:2020ani} suggest that no finite volume effects should be visible for $L_i \geq 20$ at $\beta=6.0$ where $i=x, y, z$. Hence, we expect that some finite volume effects might appear to show up for $L_x < 20$. Nevertheless, within our statistical precision no such effects appear to rise and there is a well defined plateau for the glueball ground state $0^{++}$ along the region $L_x \in [9,30]$.

As we move to smaller values of $L_x$  ($ L_c < L_x < 9 $), due to the fact that one of the spatial directions of the lattice becomes critically small, our variational calculation can capture an interacting torelon-antitorelon (di-torelon) state which, as expected, vanishes as we approach the critical length. The mass of the state is not just equal to twice the mass of the torelon but there is a non-zero contribution due to the interaction. The appearance of such a state in the glueball calculation is not prohibited by centre symmetry and therefore there is some non-zero overlap with our variational basis of operators. Of course as the number of colors $N$ increases we expect that such overlaps will be suppressed. An investigation of the glueball spectrum in the large-$N$ limit~\cite{Athenodorou_preparation} will demonstrate up to what value of $N$ such states are visible. Because the main contribution in these states comes from twice the mass of the torelon for a given $L_x$, as expected the mass of the ground state of our variational calculation decreases with $L_x$. In general these states can be considered to be finite volume states which vanish as we increase all the spatial lattice sizes. In addition, given that the phase transition is weakly first-order, close to the critical length the masses of the glueballs might be governed by the critical exponents of the theory. This means that the mass of the $0^{++}$ glueball might experience a drift towards zero as we approach the critical length from above. For larger values of $N$ where the aforementioned transition is strong first order the naive expectation is that the transition does not affect the glueball mass. To provide a definite answer on what these states are a dedicated investigation should be carried out; however this goes beyond the scope of this work.

Finally as expected, 
when we decrease $L_x < L_c$,
the calculation gives access to glueball screening masses.
 So, as a matter of fact one can think of the theory on a lattice of $L_{\tau} \times L^2 \times L_Z $ with $L_{\tau} = L_{x} < L = L_Z$ and $L_{\tau} = 1/aT$ where $T$ is the temperature. Our operators are build to project onto the ground state of $0^{++}$ for a calculation at $T=0$ and will, thus, capture this state. The mass appears to increase with temperature reflecting the large thermal shift of the $0^{++}$ mass.

Hence, our data for $\beta=6.0$ outline three different regions in $L_x$,
similar results have also been obtained for $\beta=6.2$. To summarize:
\begin{enumerate}
\item For $l_x <  l_c$ we obtain {\it glueball screening masses}.
\item For $l_c(\beta) \sqrt{\sigma} < l_x\sqrt{\sigma} \lesssim 2 $  we may obtain {\it finite volume effects  such as di-torelon and glueball states influenced by virtual glueball exchanges or the glueball mass influenced by the weakly first order phase transition}.
\item For $ 2 \lesssim l_x\sqrt{\sigma} < \infty $  we extract {\it the $0^{++}$ glueball ground state}.
\end{enumerate}

The $0^{++}$ glueball ground state, for the three values of $\beta$ considered in this work and extracted for a homogeneous spatial lattice where the octahedral subgroup of rotations is restored, are provided in Table~\ref{tab:table_tension_glueball}. These results are in good agreement with recent calculations such as Ref.~\cite{Athenodorou:2020ani}. A careful inspection over the gluball masses suggests that $m_{0^{++}}/\sqrt{\sigma}$ does not scale linearly with $a^2 \sigma$, this is due to the fact that the glueball mass for the coarsest lattice acquires scaling corrections of higher power $(a^4 \sigma^2)$, this can be seen clearly in Figure 11 of Ref.~\cite{Athenodorou:2020ani}.

\begin{figure}[htb]
  \begin{center} 
 \rotatebox{0}{\includegraphics[width=8cm]{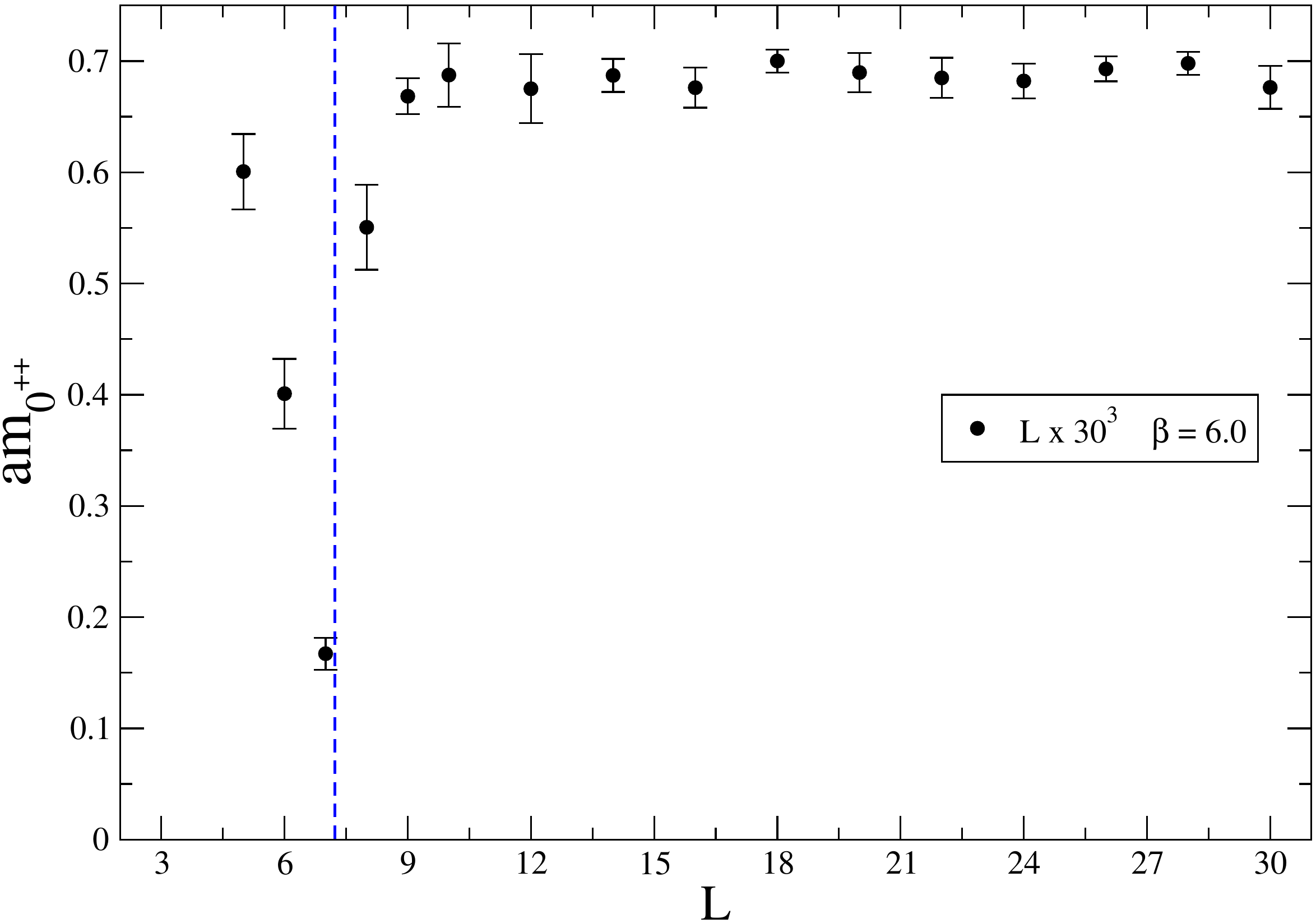}}
\caption{\label{fig:glueball_lx} The mass of the ground state of the variational calculation as a function of $L_x$ for $h=0$. The blue dashed vertical line corresponds to the critical value of length $L_c$.}
  \end{center}
\end{figure}

\subsubsection{Torelons in the non-deformed theory}
\label{sec:Torelon_non_deformed}

In this case, we calculate the ground state energy $m_T(l)$ of a flux tube of length $l$ that closes on itself by winding once around a spatial torus of size $l$. 
For a theory without a center deformation and with the torelon being formed along a direction with size larger than the deconfining length $l_c = aL_c$ its mass is given by the ground energy of a relativistic closed (non-critical) string. It has been demonstrated that a very good approximation for the ground state of the torelon is provided by the Nambu-Goto formula:
\begin{equation}
m_T(h=0,l>l_c)
\stackrel{NG}{=}
\sigma l \left(1-\frac{2\pi}{3\sigma l^2}\right)^{1/2}\,,
\label{eqn_Nambu-Goto}
\end{equation}
which derives from the light-cone quantisation of the bosonic string. Physically it arises from the regularised  sum  of  the  zero-point  energies  of  all  the  quantised  oscillators  on  the  string. It is known to provide an excellent approximation to the lattice calculations \cite{AABBMT-string} for reasons that have now become well understood (see for instance Refs.~\cite{string_theory1,string_theory2}.)


As a matter of fact one expects that this approximation works well for flux-tube lengths of $l \sqrt{\sigma} > 2.5$~\cite{AABBMT-string}. Thus, by fitting the ground state as a function of the length $l$ to this expression we can extract the string tension $\sigma$. The value of the string tension for each value of $\beta$ is listed in table~\ref{tab:table_tension_glueball}.

\begin{table}[h]
    \centering
    \begin{tabular}{|c|c|c|c|} \hline
         $\beta$ & $a^2 \sigma$ & $am_{0^{++}}$ & $L_c$ \\ \hline  \hline
         5.8 & 0.09882(79)& 0.8567(349)  & 5.000(5) \\ \hline
         6.0 & 0.04644(60) & 0.6763(192) & 7.225(19) \\ \hline
         6.2 & 0.02487(20) & 0.5218(147) & 9.892(46)  \\ \hline
    \end{tabular}
    \caption{The value of the string tension calculated using the NG formula, the value of the scalar glueball mass at $T=0$ as well as the critical length for the three values of $\beta$. The critical length has been extracted by cubic spline interpolations of the results taken from Ref.~\cite{Lucini:2003zr}.}
    \label{tab:table_tension_glueball}
\end{table}

The spectrum of the ground state as well as the low-lying spectrum of the torelon has been investigated extensively~\cite{AABBMT-string} and our results are in good agreement with previous findings. Hence, this part of our study does not provide new knowledge. Nevertheless, it provides the energy scale for each different value of $\beta$ we consider in our work as well as a comparison with the spectrum of the torelons for the trace deformed case; we will comment on that in section~\ref{sec:torelons_deformed_theory}. In Fig.~\ref{fig:torelon_undeformed} we present the ground state mass of the torelon $m_T/\sqrt{\sigma}$ as a function of its length $l \sqrt{\sigma}$ for the three values of $\beta$.
\begin{figure}[htb]
  \begin{center} 
     \rotatebox{0}{\includegraphics[width=9.5cm]{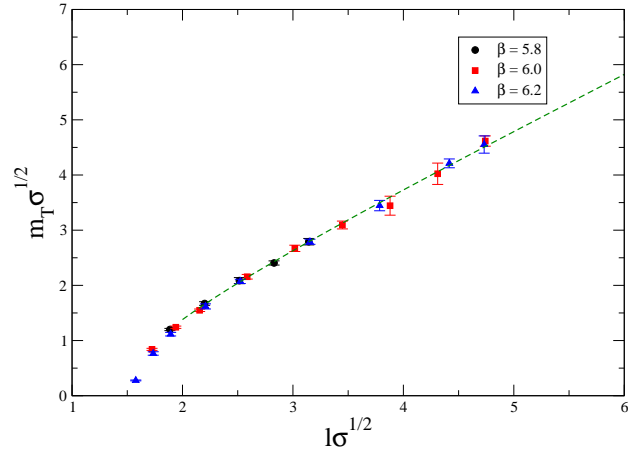}}
\caption{The mass of the torelon in units of $\sqrt{\sigma}$ as a function of the flux-tube length in dimensionless units. The line corresponds to the Nambu-Goto formula given in Eq.~(\ref{eqn_Nambu-Goto}). \label{fig:torelon_undeformed}}
  \end{center}
\end{figure} 

\subsection{Results for the trace deformed theory}
\label{sec:deformed_theory}
We turn now to the case of the trace deformed theory.
Practically, we kept $L_x < L_c$ as well as $L_y, L_z, L_t \gg L_c$, 
and we changed $h$. 
For most simulations 
we fixed $a L_x \simeq 0.54$~fm
exploring different lattice spacings, considering in particular  
$L_x= 4, \ 6, \ 8$ for $\beta=5.8, \ 6.0, \ 6.2$ 
respectively, with $L_y=L_z=L_t = 30$ in all cases. 
In addition, we performed simulations for $L_x = 6$ at 
$\beta = 6.2$ and $L_x = 8$ at $\beta = 6.1$ in order to explore
different values of the compactification length, 
respectively $l_x \simeq 0.4$~fm and $l_x \simeq 0.62$~fm.

Measurements of glueball
and torelon states were taken along each of the homogeneous directions,
in order to increase statistics, using 
around 4000 decorrelated configurations for each value of $h$.

\subsubsection{Glueballs in the trace deformed theory}
\label{sec:glueballs_deformed_theory}
Similarly to the case of the non-deformed theory, we performed a measurement 
of the glueball scalar mass using the variational technique with a basis of 
operators built out of the simple plaquette as well as the rectangular 
operators, using smearing and blocking techniques in order 
to enhance the overlap onto the physical states. We extracted the absolute 
ground state of the glueball mass which corresponds to the $0^{++}$ ground 
state as a function of $h$ for three values of the lattice spacing 
i.e.~three values of $\beta$. Results obtained for 
$l_x \simeq 0.54$~fm are shown in Fig~\ref{fig:glueball_h}
where we also report, for comparison, the expectation 
value of the Polyakov loop, which becomes zero once the center symmetry is 
restored.

\begin{figure}[htbp]
  \begin{center} 
    \rotatebox{0}{\includegraphics[width=8.7cm]{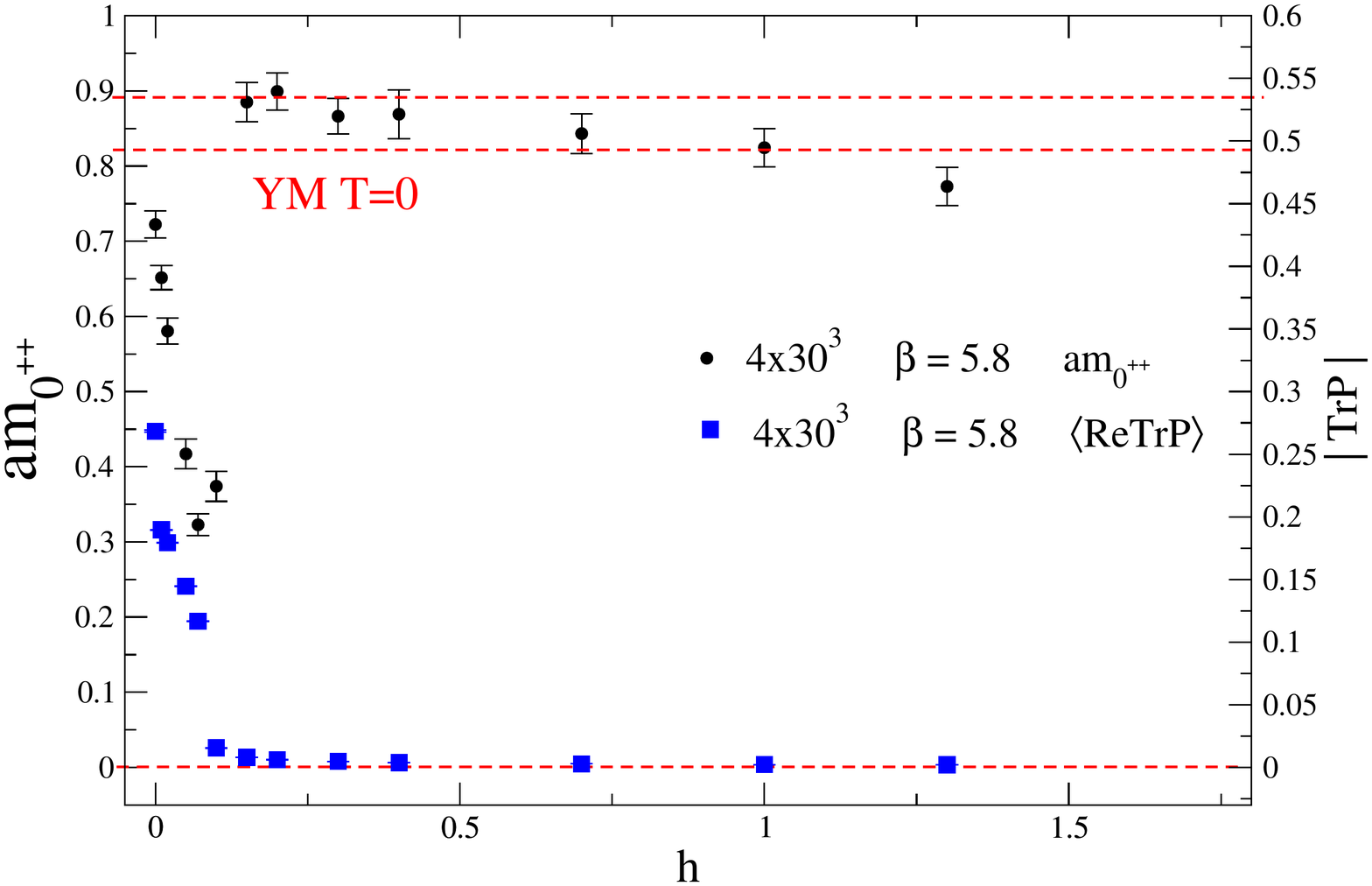}}
    \rotatebox{0}{\includegraphics[width=8.7cm]{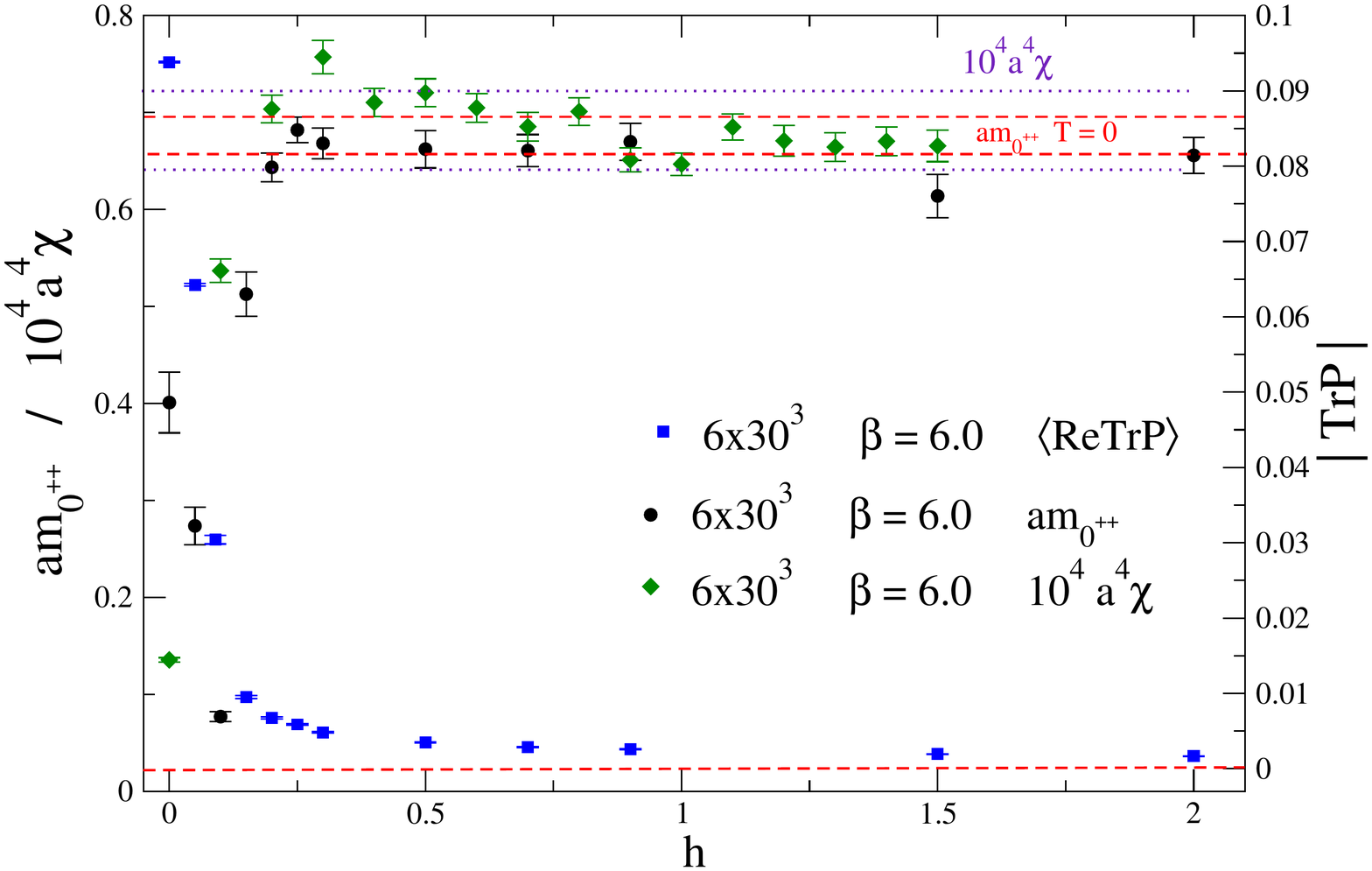}}
    \rotatebox{0}{\includegraphics[width=8.7cm]{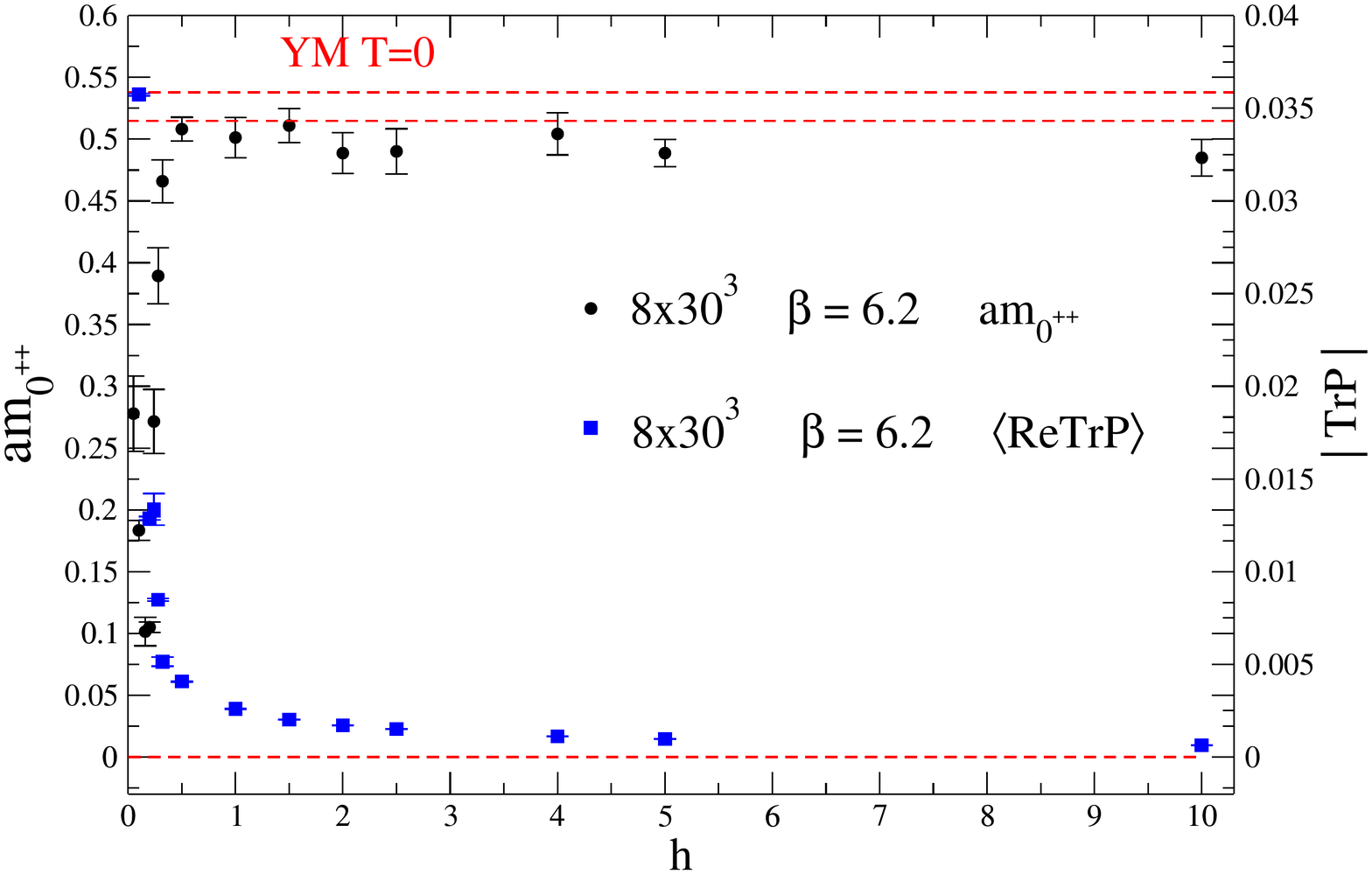}}
\caption{\label{fig:glueball_h} The mass of the ground state resulting out of the variational calculation for glueballs (in black) as well as the average value of the Polyakov loop (in blue) in the trace deformed theory as a function of $h$ and the three values of $\beta$. For $\beta = 6.0$ we also report,
for comparison, results for the topological susceptibility, which 
have been taken from Ref.~\cite{Bonati:2018rfg}.}
  \end{center}
\end{figure}

The plots resemble to an adequate extent the behaviour of the glueball mass 
observed for the case of the non-deformed theory, where we alter the lattice 
size along the $x$ direction. For very small values of $h$ 
center symmetry is still broken, so that the 
variational calculation captures 
the screening mass. 
As we keep increasing $h$, center symmetry gets restored and the theory 
experiences a transition to confinement, so that the correlators
start capturing confining states such as bound states 
of two closed flux-tubes or glueball states: the resulting 
glueball mass shows a dip after which, 
deep in the confined phase, it reaches a well defined plateau.
This plateau appears to be consistent with the glueball mass extracted 
at $T=0$. 
The critical value of $h$, as 
expected, depends on $\beta$, i.e.~on the lattice spacing. 
Analogous results have been obtained for the other 
explored values of $l_x$: the plateau values in the re-confined phase 
are summarized in Table~\ref{tab:table_plateau_am0++}.

\begin{table}[h]
    \centering
    \begin{tabular}{|c|c|c|c|} \hline
         $\beta$ & $L_x$ & $l_x \ \mathrm{[fm]}$ & $am_{0^{++}}$ \\ \hline  \hline
         5.8 & 4 & 0.54  & 0.85(4) \\ \hline
         6.0 & 6 & 0.56  & 0.66(2) \\ \hline
         6.2 & 6 & 0.40  & 0.46(2)  \\ \hline
         6.1 & 8 & 0.62  & 0.50(2)  \\ \hline
         6.2 & 8 & 0.54  & 0.59(2)  \\ \hline
    \end{tabular}
    \caption{The plateau value of $am_{0^{++}}$ for the different lattice setups used.}
    \label{tab:table_plateau_am0++}
\end{table}

It is interesting to notice that this behavior resembles closely that 
of the topological susceptibility $\chi$,
which is reported for comparison in one case ($\beta = 6.0$)
and reaches a plateau at approximately the same values of $h$:
a similar behavior is observed for the other cases.
That shows
that the re-confined compactified theory recovers most of the 
non-perturbative features of the original $T = 0$ theory at the 
same time.

Just for $\beta = 6.2$ we report results extended to larger values 
of the deformation parameter, namely up to $h=10$, showing 
that the glueball mass in this case is also quite stable:
{this is an interesting aspect, however it is not
essential and indeed the plateau is not so stable for other values
of the lattice spacing. In general, we are more interested in 
the small $h$ region, where the properties of the 
re-confined phase can be compared to those of the standard
confined phase more closely, trying also to determine
a sort of $h$-dependent effective compactification radius, 
as we illustrate in more details in the next paragraph.}



\subsubsection{Trying to match the $h$-dependence of the deformed theory to the $l$-dependence of the undeformed theory} 

The close resemblance between the dependence of the glueball mass on $l$ at $h = 0$, and of that 
on $h$ at fixed $l$, suggests a possible interpretation of a non-zero $h$ 
in terms of an effective, $h$-dependent compactification size $l_{\rm eff}(l,h)$, i.e.~a match of 
results at fixed $l$ and $h \neq 0$ to those obtained with a compactification size $l_{\rm eff}$ 
at $h = 0$. In order to further
test this possibility, we have worked on a possible ansatz for such dependence.

Since a positive/negative $h$ will favor/disfavor confinement, this is 
analogous to increasing or decreasing $l$. Hence, the dependence of such
$l_{\rm eff}$ on $h$ must be odd at least at the lowest order in a Taylor expansion around
$h = 0$. On the other hand, it is sensible to look for a description in
terms of $l_{\rm eff}$ only for small values of $h$, since large values are expected to modify 
the theory more substantially.

Therefore, we have considered the following linear ansatz
\begin{equation}
l_{\rm eff} (l,h) = l\ ( 1 + A h)\,,
\label{effectivel}
\end{equation}
where $A$ is a dimensionless parameter, which in general is expected
to depend on the ultraviolet (UV) cut-off since $h$ is a bare parameter.

Such ansatz turns out to work reasonably well. In Fig.~\ref{fig:glueball_effective}
we compare results obtained for $\beta = 6.0$, either at $h = 0$ and variable
$L_x$ or at fixed $L_x = 6$ and variable $h$, showing that a reasonable match is obtained by taking 
$A \simeq 2.10$, with an uncertainty estimated around 10\% based on a by eye matching.
In Fig.~\ref{fig:glueball_effective}
we report additionally also results obtained for $\beta = 6.2$ and $L_x  = 8$, just to 
show that in this case $A \simeq 1.35$, confirming that it is indeed cut-off dependent.

\begin{figure}[t!]
  \begin{center} 
    \rotatebox{0}{\includegraphics[width=9.5cm]{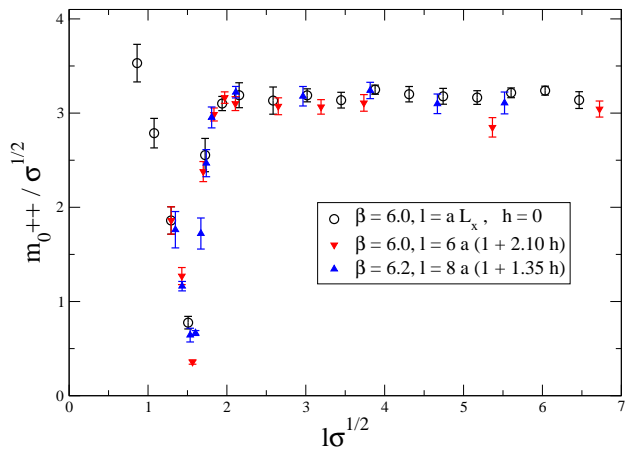}}
    \caption{\label{fig:glueball_effective} Gueball masses obtained either 
at $h = 0$ and variable $l$ or at fixed $L_x$ and variable $h$. In the latter
case the horizontal scale has been fixed according to the ansatz 
in Eq.~(\ref{effectivel}), where the constant $A$ has been tuned 
based on a by eye matching.}
  \end{center}
\end{figure}

From Fig.~\ref{fig:glueball_effective} one could get the wrong suggestion 
that the ansatz works well even for large values of $h$. However one should 
consider that, once in the confined phase, the glueball mass is in fact 
$l$-independent (hence volume-independent), even in the undeformed theory.
A better feeling about the range of validity of the ansatz in Eq.~(\ref{effectivel}) 
is therefore obtained by looking at the torelon mass, which turns out to have 
a non-trivial dependence on $l$ in the confined phase.

\subsubsection{Torelons in the trace deformed theory}
\label{sec:torelons_deformed_theory}

In order to discuss the behavior of the torelon ground energy 
as a function of the deformation parameter, we start by 
illustrating the results obtained at 
$\beta = 6.0$ and $L_x = 6$, which are reported in 
Fig.~\ref{fig:torelon_effective} and compared with results 
obtained in the undeformed theory, adopting the same matching 
in terms of an effective compactification size fixed 
from the glueball mass.



\begin{figure}[t!]
  \begin{center} 
    \rotatebox{0}{\includegraphics[width=9.5cm]{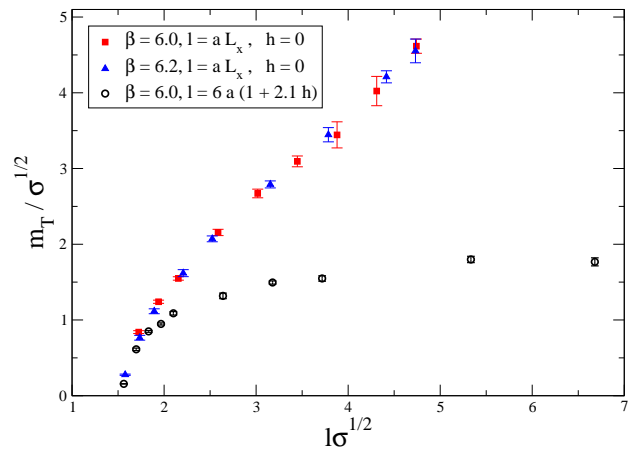}}
    \caption{\label{fig:torelon_effective} A comparison similar
to that of Fig.~\ref{fig:torelon_effective}, performed in this case
for the torelon masses. The rescaling parameter $A$ has been kept
fixed to the value already matched for the glueball masses, in order
to see if the ansatz in Eq.~(\ref{effectivel}) works well for all 
observables, at least for small values of $h$.}
  \end{center}
\end{figure}

It is clear that the matching works 
pretty well also for torelon masses as long as $h$ is small
(in particular for $h \lesssim 0.2$). Then deviations are larger 
and larger and, contrary to the linearly rising behavior observed 
in the undeformed theory, results seem to reach a well defined plateau
at asymptotically large values of $h$. A similar behavior is 
observed also in other cases, as an example in 
Fig.~\ref{fig:torelon_h} we show the ground mass of the torelon as a function of 
$h$ for different values of the lattice spacing and $l \simeq 0.54$~fm.
Assuming
that the ground energy indeed reaches a plateau with $h$,
we have tried to fit data in Fig.~\ref{fig:torelon_h} according to the following ansatz
\begin{eqnarray}
  am_{T}(h) = am_{T}(\infty) + b e^{-c h} \, .
  \label{eqn:ansatz}
\end{eqnarray}
The fit works well, i.e.~with a $\chi^2/\textrm{d.o.f.}$ of order $\sim 1$,
if data at the lowest values of $h$ are discarded.

It is interesting to ask what the plateau values should be compared with.
Since large values of $h$ tend to suppress local fluctuations of the 
Polyakov loop more and more, it is reasonable to look for predictions 
obtained in the semiclassical regime, such as those reported in Ref.~\cite{Unsal:2008ch}.
When we switch on the trace deformation along a circle of small radius, that is to say $l_x \ll 1 / \sqrt{\sigma}$, the trace potential in the action becomes minimum if the Polyakov loop along this circle acquires a diagonal value of $P = {\rm Diag} \left( 1, e^{2 \pi i / N}, e^{4 \pi i / N}, \dots,  e^{4 \pi i (N-1)/ N}   \right)$ up to conjugation by an arbitrary $SU(N)$ matrix. If one works in a gauge in which the Polyakov loop $P$ is diagonal, and  using  gauge-dependent  language,  this  configuration  may  be  regarded  as breaking the gauge symmetry down to the maximal Abelian subgroup i.e. $SU(N) \to U(1)^{(N-1)}$.

The modes of the diagonal components of the $SU(N)$ gauge field with no relative momentum along the compactified direction describe photons associated with the $SU(N)$ Cartan subgroup while modes with non-zero relative momentum form a Kaluza-Klein tower with masses being integer multiples of $2 \pi /L$~\cite{Unsal:2008ch}: 
\begin{eqnarray}
  m_T(l) = \frac{2 \pi}{ N l} = \frac{2 \pi}{ 3 l}.
  \label{eq:deformed_tube}
\end{eqnarray}
Such prediction should work in principle
for $N \Lambda l \ll 1$, where $\Lambda$ is the QCD scale, 
which is not the case for our simulations. However,
as a matter of fact, the results of Fig.~\ref{fig:torelon_h} show that  
the plateau values are quite close to the $2 \pi / 3 l$
prediction, even if 
$l$ is only slightly smaller than 
the deconfining length $l_c$.
One should consider that, in this case, we are 
discussing a sort of large-deformation limit of what is found, and 
this could have some influence on the final findings.


\begin{figure}[h!]
  \begin{center} 
    \rotatebox{0}{\includegraphics[width=8cm]{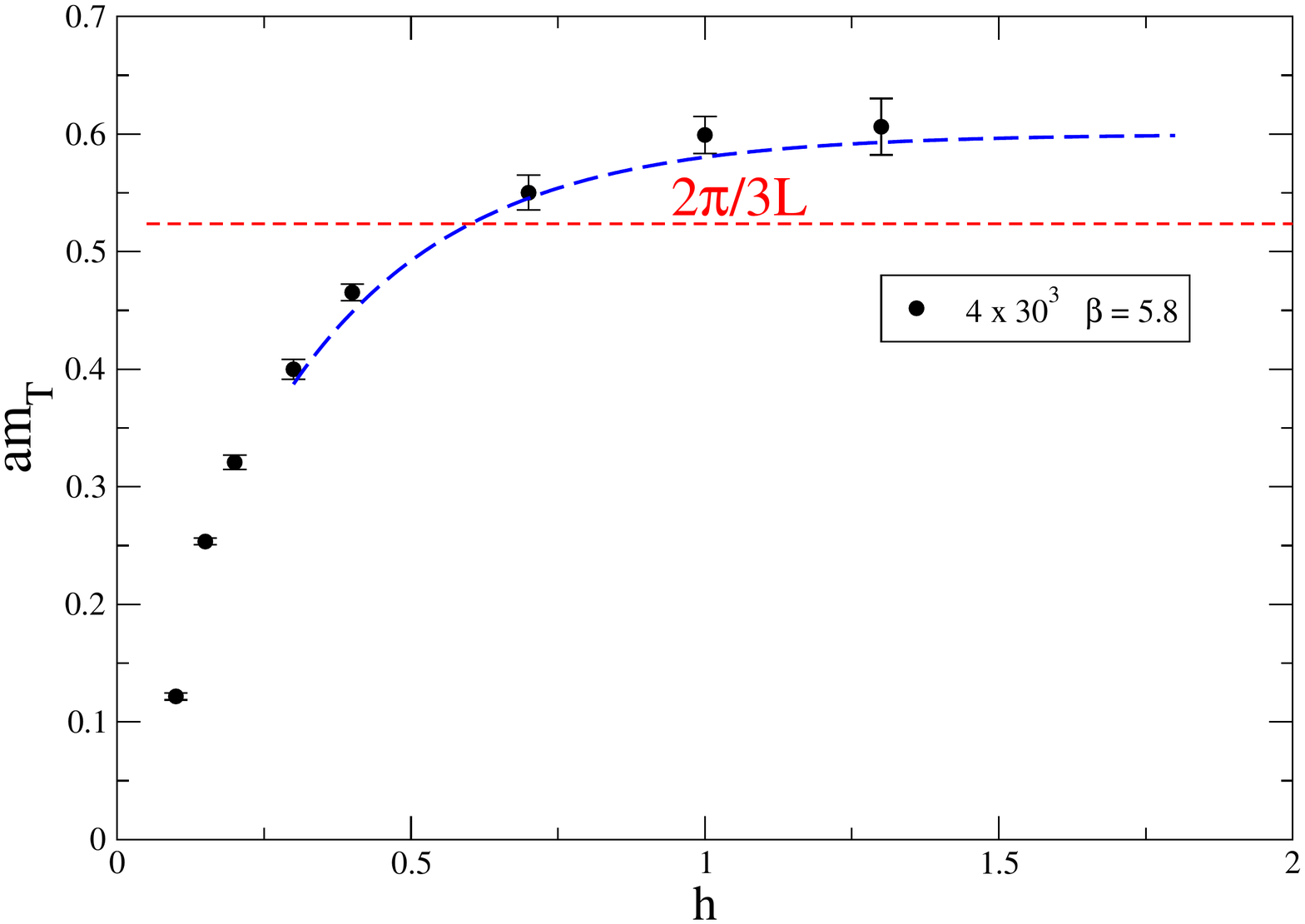}}
    \rotatebox{0}{\includegraphics[width=8cm]{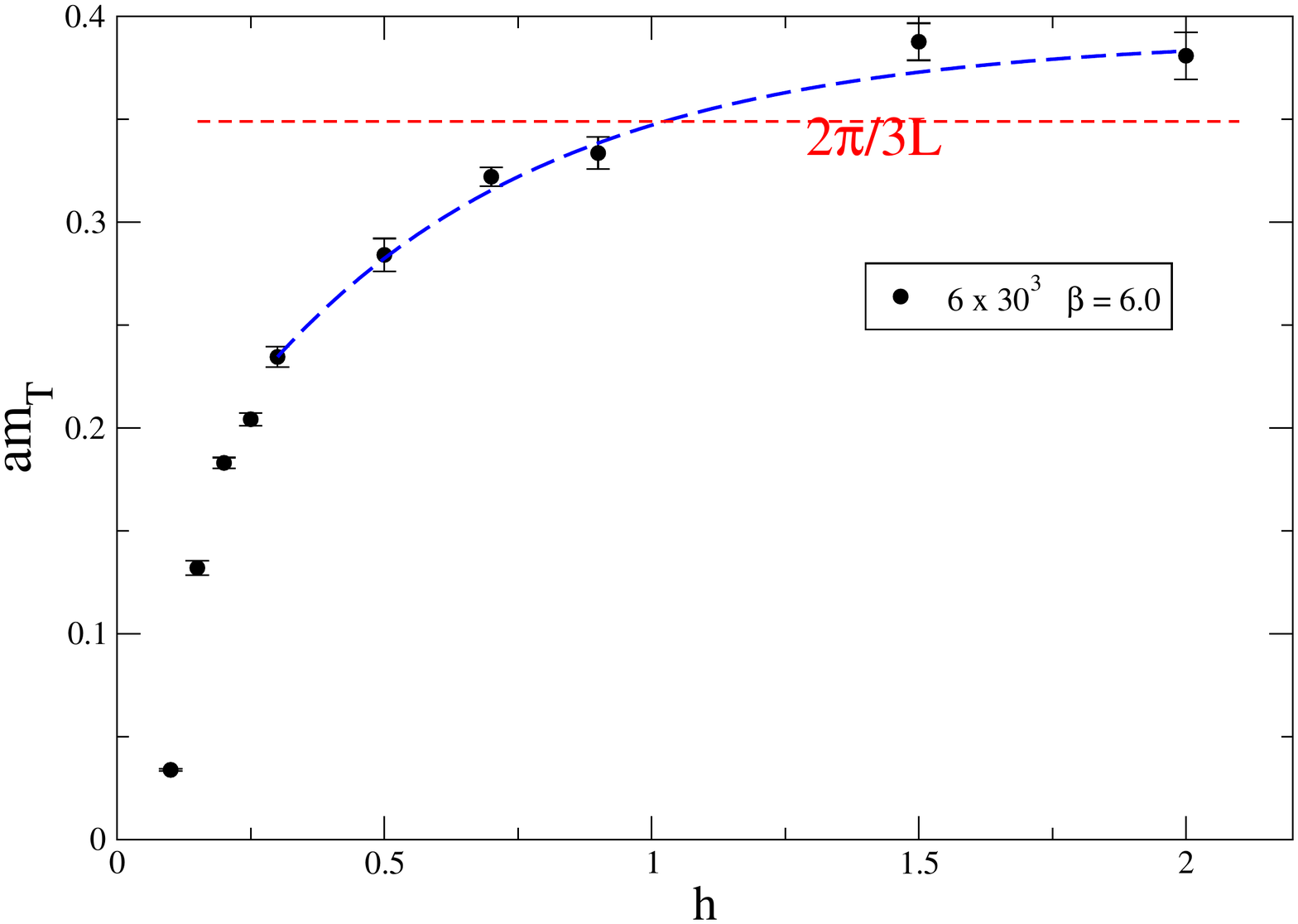}}
    \rotatebox{0}{\includegraphics[width=8cm]{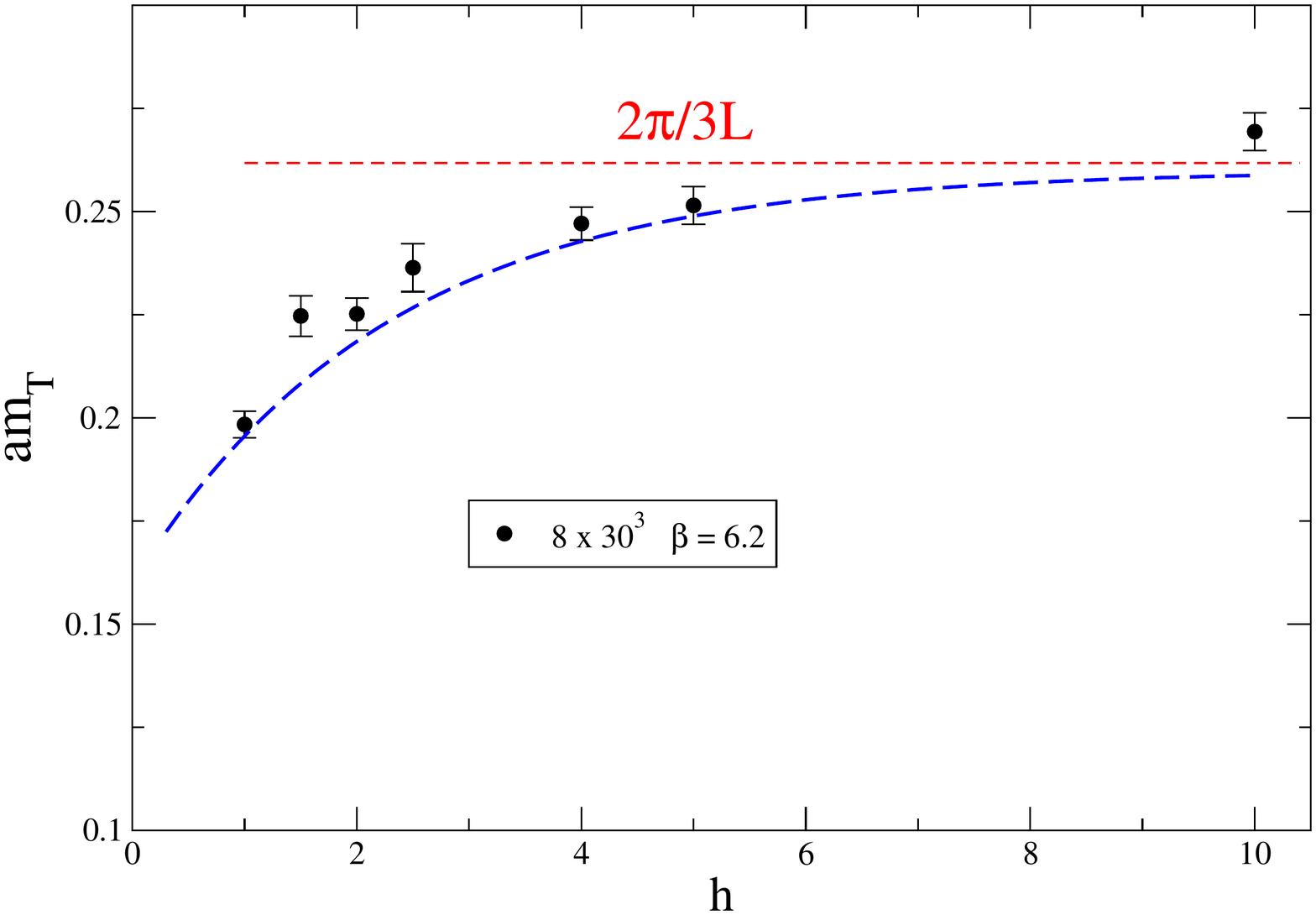}}
\caption{\label{fig:torelon_h} The mass of the ground state resulting out of the variational calculation for torelons in the trace deformed theory as a function of $h$ and the three values of $\beta$.}
  \end{center}
\end{figure}

In 
Fig.~\ref{fig:torelon_continuum} we report the three
plateau values of Fig.~\ref{fig:torelon_h} 
for $a m_{T} L$ as a function of $1/L^2$. 
Mind that $1/L^2 = a^2/l^2$ with $l$ being fixed in physical units, 
thus this plot corresponds to an extrapolation of $a m_{T} L$ to the 
continuum limit and shows two interesting things at the same time. 
First, the three points can be adequately fitted 
with a straight line,
i.e.~assuming that UV corrections are $O(a^2)$, so that
the plateau value corresponds to a well defined continuum
quantity which we derive to be $a m_{T} L = 2.1(2)$. 
Second, this number is in good agreement with the theoretical expectation 
in the semiclassical regime, i.e.~$2 \pi/3 \simeq 2.094$.

\begin{figure}[t!]
  \begin{center} 
    \rotatebox{0}{\includegraphics[width=9.5cm]{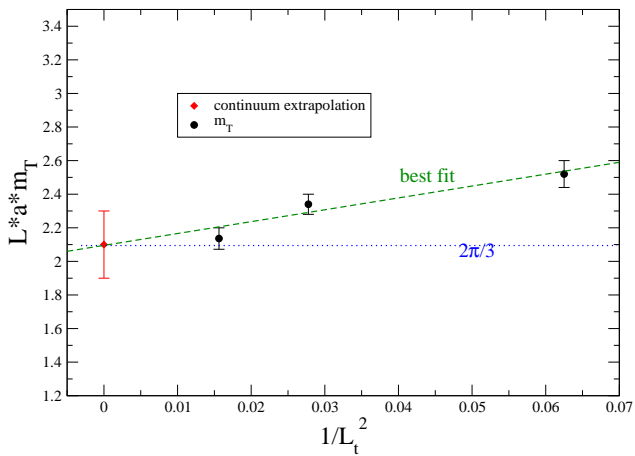}}
    \caption{\label{fig:torelon_continuum} The continuum extrapolation of the torelon masses extracted by fitting the ansatz of Eq.~\ref{eqn:ansatz} on the torelon data for the three values of $\beta$. The dotted line represents the theoretical prediction $m_T=2 \pi/3$~\cite{Unsal:2008ch}, while the 
dashed line represents the results of a continuum extrapolation assuming
$O(a^2)$ corrections, which yields a $\chi^2/{\rm d.o.f.} = 1.7/1$.}
  \end{center}
\end{figure}

However, in order to test if this is just a fortuitous 
coincidence, we have decided to repeat our study for two different 
values of $l$, one smaller and one larger.
In particular, we explored the case $L_x = 6$ at 
$\beta = 6.2$, corresponding to $l \sim 0.4$~fm, and the 
case $L_x = 8$ at $\beta = 6.1$, corresponding to $l \sim 0.62$~fm.
It is not easy, in particular, to go to smaller values,
since it is difficult to keep lattice artifacts 
well under control as $l$ is reduced with a fixed amount of 
lattice spacings $L_x$ in the compactified direction 
(see Ref.~\cite{marco:lattice} for a discussion on this point).

\begin{figure}[t!]
  \begin{center} 
    \rotatebox{0}{\includegraphics[width=9.5cm]{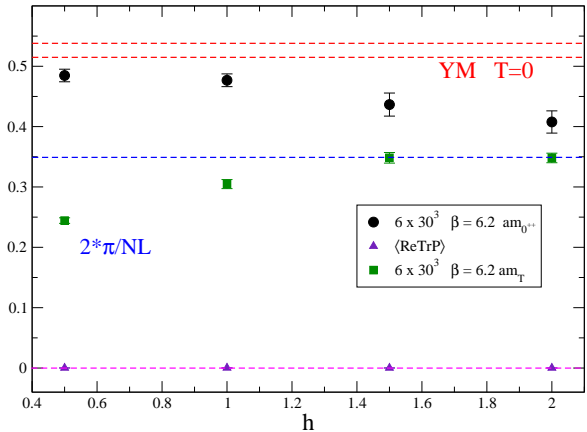}}
    \caption{Glueball and torelon masses, as a function of $h$, for 
$L_x = 6$ and $\beta = 6.2$, corresponding to $l \simeq 0.4$~fm. 
We report also values obtained for the Polyakov loop expectation value,
showing that center symmetry is restored for all explored values of $h$.}
\label{fig:glueball_Nt6}
  \end{center}
\end{figure}

In Fig.~\ref{fig:glueball_Nt6} we report, as 
a function of $h$, values obtained for $l \sim 0.4$~fm
for both the glueball and the torelon mass.
One can see that the plateau region for the glueball mass is somewhat reduced
in this case (we consider the region up to $h \simeq 1$), while
the large-$h$ plateau for the torelon mass is still well defined.
The plateau values for these new values of $l$ are reported in 
Fig.~\ref{fig:torelon_glueball_Nt6}, where they are compared
with those previously obtained for $l \simeq 0.54$~fm, with results
obtained for the glueball masses and with the exploratory results
obtained for $SU(4)$ which are discussed in the next Section.
One can appreciate that the glueball mass does not show, within errors, 
a significant dependence on $l$, while the torelon mass 
has a well defined dependence which is still compatible with 
the $2 \pi /(3l)$ prediction, apart from the point at the 
largest explored value of $l$. Of course it would be interesting,
in future studies, to extend this investigation to smaller
values of $l$, possibly to the point where the torelon 
mass and the glueball mass cross each other.

\begin{figure}[t!]
  \begin{center} 
    \rotatebox{0}{\includegraphics[width=9.5cm]{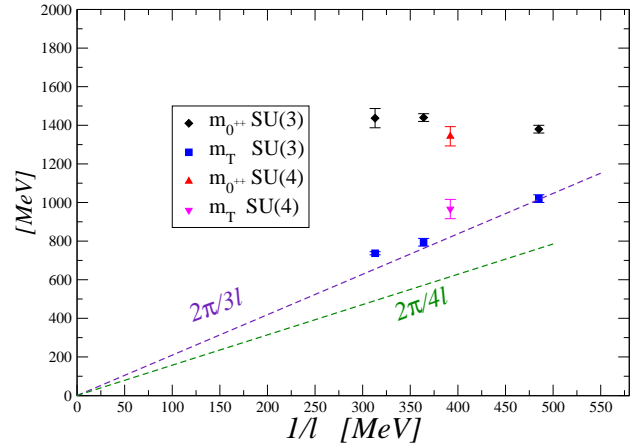}}
    \caption{Glueball and torelon masses as a function of $1/l$ for both 
$SU(3)$ and $SU(4)$.
The dashed lines represent the prediction from Ref.~\cite{Unsal:2008ch}.}
\label{fig:torelon_glueball_Nt6}
  \end{center}
\end{figure}

\section{SU(4)}

\label{sec:su4}

The extension of our study to $SU(4)$ in this context is exploratory 
and aimed at enlightening the main differences emerging when moving to
$N > 3$; in particular, we expect to go even farther from the 
semiclassical regime dictated by $N \Lambda l \ll 1$, so that,
in principle, the matching of torelon masses to the prediction of Eq.~(\ref{eq:deformed_tube})
should be worse.

As explained in Section~\ref{sec:setup}, in this case one can introduce two different
deformation parameters, $h_1$ and $h_2$, leading to a more structured phase diagram,
which has been sketched in Ref.~\cite{Bonati:2019kmf} and is characterized by
different phases where center symmetry is completely broken, partially broken, or completely 
restored. In this study we consider 
only the case of the diagonal deformation $h = h_1 = h_2$, which guarantees 
a complete restoration of center symmetry for large enough $h$ and 
is thus adequate to our purposes.

\begin{figure}[t!]
  \begin{center} 
    \rotatebox{0}{\includegraphics[width=9.5cm]{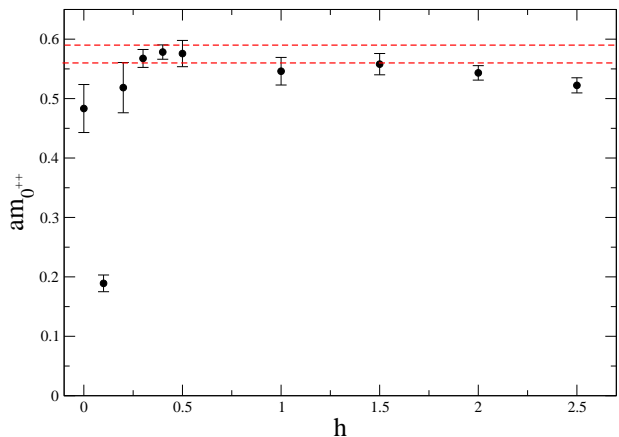}}
    \caption{Glueball mass for $\beta$ = 11.15 on a 6$\times \ 32^3$ lattice with both the deformations switched on,
$h_1 = h_2 = h$.}
\label{fig:glueball_mass_su4}
  \end{center}
\end{figure}

\begin{figure}[t!]
  \begin{center} 
    \rotatebox{0}{\includegraphics[width=9.5cm]{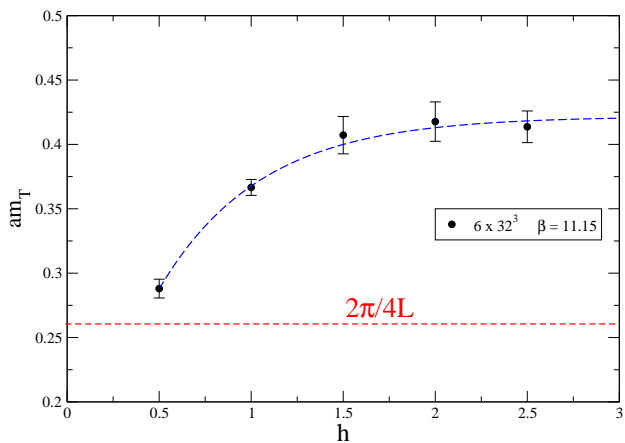}}
    \caption{Torelon mass for $\beta$ = 11.15 on a 6$\times \ 32^3$ lattice as a function of $h = h_1 = h_2$.}
\label{fig:torelon_mass_su4}
  \end{center}
\end{figure}

We have also considered a single value of the compactification size,
$l \simeq 0.50 \ \mathrm{fm}$ corresponding to a deconfined phase
in the undeformed theory, at a fixed value of the UV cut-off,
$a\simeq 0.083 \ \mathrm{fm}$, with no aim towards an extensive 
study of $l$-dependence or continuum extrapolation. Results for the glueball
mass as a function of $h$ are presented in Fig.~\ref{fig:glueball_mass_su4}:
the behavior resembles closely that already observed for $SU(3)$, reaching 
a well defined plateau for $h \gtrsim 0.5$, which corresponds to a phase 
where center symmetry is completely restored~\cite{Bonati:2019kmf}. Moreover,
the plateau value is in good agreement with the $T = 0$ value,
which is reported in the figure as well. The $T=0$ result has been computed on an isotropic lattice with volume of $30^4$, with the same value of $\beta$ and it is consistent with results from Ref.~\cite{Athenodorou:2021qvs}.

Also results for the torelon ground state, which are presented in 
Fig.~\ref{fig:torelon_mass_su4}, show a behavior quite similar to that
observed in $SU(3)$, with an approach to a plateau value at large values
of $h$ which also in this case has been fitted using the ansatz reported
in Eq.~(\ref{eqn:ansatz}). However, in this case the plateau value turns
out to be significantly larger than the semiclassical prediction of Eq.~(\ref{eq:deformed_tube}):
this is more clear from Fig.~\ref{fig:torelon_glueball_Nt6}, where the results
obtained in $SU(4)$ for both the glueball mass and the torelon mass
are compared with the $SU(3)$ results.

\section{Discussion and Conclusions}
\label{sec:conclusion}

The purpose of this investigation was that of exploring the properties of 
trace deformed Yang-Mills theories from the point of view of their physical spectrum.
Such theories are defined on a space-time with at least one
small compactified spatial direction and in the presence of trace deformations
which prevent the breaking of center symmetry, giving the possibility to check
the conjecture of volume independence for $SU(N)$ gauge theories in the large-$N$ limit.
Lattice simulations have already provided successful confirmations at moderate
values of $N = 3,4$ for what concerns the topological properties 
of the theory~\cite{Bonati:2018rfg,Bonati:2019kmf}. 

Extending the investigation to the physical states of the theory
permits to better clarify the relation with the standard large-volume
theory, by distinguishing states in the deformed theory which have a clear
correspondence with the standard physical states of the original theory,
and states which are linked to the large energy scale associated with the
small compactified direction  
and are expected to decouple as the compactification radius 
goes to zero.
Having this in mind, we have performed a numerical study 
of trace deformed $SU(3)$ and $SU(4)$ gauge theories, focussing on two kinds of physical states:
scalar glueballs and torelons defined around the compactified direction.
The investigation has been performed for different values
of the compactification size and, just for one case, using also different values of the lattice spacing,
in order to check the absence of significant cutoff effects and perform a
continuum extrapolation.

The study of scalar glueballs has fully confirmed what already observed
for the case of $\theta$-dependence, i.e.~a striking quantitative agreement
with the values of the original large volume theory, which is observed as soon
as the deformation is strong enough to make center symmetry unbroken.
This is a further confirmation of volume independence, already observed even for 
moderate values of $N$ in the case of topological observables.

Further insight has been achieved by a comparison
of results for the glueball mass obtained at fixed compactification size $l$ as a function
of the deformation parameter $h$, with results obtained in the undeformed theory 
as a function of $l$. In particular, we have shown that results in the deformed
theory can be interpreted in terms of an $h$-dependent effective 
compactification size, $l_{\rm eff}(l,h)$, see in particular Eq.~(\ref{effectivel})
and Fig.~\ref{fig:glueball_effective}.

Torelons have shown a different behavior. The matching with results from 
the undeformed theory in terms of an effective compactification size works
well only for small values of $h$, see Fig.~\ref{fig:torelon_effective}. 
The different behavior, with respect to glueballs, can be easily understood 
in terms of the fact that the torelon mass is not volume independent even in 
the confined phase, where it increases linearly with $l$, 
see Fig.~\ref{fig:torelon_continuum}. 

The last consideration sheds also light 
on striking realization of volume independence that is observed,
both for glueballs and for topological observables, even for moderate values 
of $N$. Such quantities are already practically volume independent in the underformed theory,
as long as center symmetry is safe, in the sense that they are in fact
temperature independent till the deconfinement temperature is approached
from below. 

On the other hand, we have shown that the torelon mass approaches a well
defined plateau value in the large $h$ limit, which moreover has a well defined 
continuum limit. We have compared such value with the available semiclassical 
prediction~\cite{Unsal:2010qh,Unsal:2008ch} for the lowest mass 
of the tower of Kaluza-Klein modes associated with the compactified direction, 
which is $2 \pi / (N l)$, hence 
$2 \pi / 3 l$ for $SU(3)$, where $l$ is the length of the compactified direction
in physical units. While the $1/l$ factor is a generic expectation for 
Kaluza-Klein modes, the additional $1/N$ factor is directly related to the way in which 
center symmetry is expected to be restored, in particular to the fact that the holonomy 
eigenvalues are evenly distributed around the complex unit circle and 
to the $SU(N) \to U(1)^N$ Higgsing of the theory. 
Such prediction is expected to work only when $N \Lambda l \ll 1$, where 
$\Lambda$ is the strong interaction scale, meaning $l \ll 1/N$ in fermi 
units. Nevertheless, we have observed a reasonable agreement 
with $SU(3)$ results even for $l$ as large as $0.4 - 0.5$~fm, while some 
deviations are visibile for $l \gtrsim 0.6$~fm; significant deviations instead 
emerge going to larger values of $N$, in particular for the $N = 4$ and $l \sim 0.5$~fm
case explored in this study.

To summarize, our numerical study provides, on one hand, a solid support to predictions 
regarding the way in which center symmetry is restored in the trace deformed theory.
On the other hand, it fully confirms that, independently of the particular way the 
restoration is achieved, the small volume unbroken theory is indistinguishable 
from the original large volume theory for a large set of physical observables, going from 
$\theta$-dependence to the spectrum of glueball states. Future studies could further 
extend such evidence by considering smaller values of $l$, larger values of $N$ 
and different relevant observables.

\acknowledgments
We would like to thank Mithat Unsal, Aleksey Cherman and Mike Teper for fruitful discussions. AA has been financially supported by the European Union's Horizon 2020 research and innovation programme ``Tips in SCQFT'' under the Marie Sk\l odowska-Curie grant agreement No. 791122. Numerical simulations have been performed at the Scientific Computing Center at INFN-PISA and on the MARCONI machine at CINECA, based on the agreement between INFN and CINECA (under projects INF19\_npqcd and INF20\_npqcd).

\end{document}